\documentclass[sigconf]{acmart}

\AtBeginDocument{%
  }

\setcopyright{acmlicensed}
\copyrightyear{2026}
\acmYear{2026}
\acmDOI{XXXXXXX.XXXXXXX}

\copyrightyear{2026}
\acmYear{2026}
\setcopyright{cc}
\setcctype{by}
\acmConference[CHI '26]{Proceedings of the 2026 CHI Conference on Human Factors in Computing Systems}{April 13--17, 2026}{Barcelona, Spain}
\acmBooktitle{Proceedings of the 2026 CHI Conference on Human Factors in Computing Systems (CHI '26), April 13--17, 2026, Barcelona, Spain}
\acmPrice{}
\acmDOI{10.1145/3772318.3790962}
\acmISBN{979-8-4007-2278-3/2026/04}

\newcommand{\newedits}[1]{\textcolor{black}{#1}}

\newcommand{\finalrev}[1]{\textcolor{black}{#1}}

\usepackage{tikz}
\usepackage{arydshln}
\usepackage{graphicx}
\usetikzlibrary{shapes.geometric, arrows}

\tikzstyle{process} = [rectangle, rounded corners, draw=black, fill=blue!10, minimum width=5cm, minimum height=1cm, text centered]
\tikzstyle{arrow} = [thick,->,>=stealth]

\begin{document}

\title[Data Repair]{Data Repair}

\author{ATM Mizanur Rahman}
\affiliation{
  \department{Computer Science}
  \institution{University of Illinois Urbana-Champaign}
  \city{Champaign}
  \state{Illinois}
  \country{USA}}
\email{amr12@illinois.edu}

\author{Syed Ishtiaque Ahmed}
\affiliation{
  \department{Computer Science}
  \institution{University of Toronto}
  \city{Toronto}
  \state{Ontario}
  \country{Canada}}
\email{ishtiaque@cs.toronto.edu}

\author{Sharifa Sultana}
\affiliation{
  \department{Computer Science}
  \institution{University of Illinois Urbana-Champaign}
  \city{Champaign}
  \state{Illinois}
  \country{USA}}
\email{sharifas@illinois.edu}

\renewcommand{\shortauthors}{Rahman et al.}

\begin{abstract}
This paper investigates data repair practices through a six-month-long ethnographic study in Bangladesh. Our interviews and field observations with data repairers and related stakeholders found that, alongside the scarcity of high-precision machinery and access to advanced software, data repair work is constrained by cross-language learning resources and the protective nature of documenting, curating, and sharing the experiences and knowledge among local peers. Repairers turning to external resources such as foreign forums and LLMs also revealed their frustrating experiences and the postcolonial ethical tensions they encountered. We noted that both anticipated technical labor and the emotionality of data were taken into account for pricing the data repair job, which contributed to their market sustainability strategies. Engaging with repair, infrastructure, and data poverty discourse, we argue that data repair practices represent a crucial challenge and opportunity for HCI in advancing global efforts toward data equity.
\end{abstract}




\maketitle

\section{Introduction}
Data repair refers to the efforts people make to regain access to lost or inaccessible datafied materials. Data Repair practices include recovering files from failing media, restoring accounts after forgotten passwords or hacks, repairing corrupted or low-resolution images, restoring images and documents from damaged paper copies, and converting formats so that important videos, audios, and documents remain usable. Although repair has long been a major theme in HCI, most of this research has focused on hardware and device maintenance \cite{jackson2014breakdown, ahmed2015learning, houston2016values}. In contrast, the field of data repair remains highly underexplored, particularly in the Global South. Understanding Global South people's data repair practices is essential for HCI and data-driven system design because people's inability to preserve and maintain their data that represents them would result in our failure in designing appropriate tools and technologies that reflect their values and needs. \newedits{While a failure of this kind risks creating data voids and exacerbating data poverty, \finalrev{leading to biased and culturally non-resonating AI systems} \cite{leidig2015quantifying, jackson201411}, the lack of HCI's understanding of such risks exposes a concerning knowledge gap of the scholarship. This persistent gap creates profound epistemic blind spots, reinforcing structural inequalities in whose data, practices, and futures are rendered visible within the knowledge infrastructures that shape HCI and global technological development.} 

Related recovery scholarship is primarily dominated by works on account recovery, password resets, and concerns around security and privacy \cite{bonneau2015secrets, schechter2009s, gerlitz2023adventures}. Our concerns of data repair, in addition, take into account both technical and social sides of salvaging information from broken phones, damaged memory cards, or cracked screens; restoring access to social and other platform accounts; format conversion of digital files; and image and document retrieval from damaged paper copies. Understanding how knowledge is produced, shared, and curated, as well as the struggles repairers face with resource constraints, ethical ambiguities, and the need to tailor services to their clients' socio-economic needs and values is essential to obtain a holistic view of this ecosystem. \newedits{For developing appropriate tech-market strategies and economic policies at the local and global levels, such insights are crucial, but currently they exist in fragmented and epistemically thin forms, an incompleteness that introduces significant epistemic opacity into decision-making processes in policy making, tech-industry practices, and HCI scholarship.}


\newedits{Our six-month-long ethnographic study addresses this knowledge gap by examining the data repair ecosystem in Dhaka, Bangladesh.} Employing interviews and field observation, we documented the venues and people of this ecosystem, the most common forms of data repair requisitions, repairers' skills and practices, and different pain points. Our research questions were: 

\begin{quote}
\textit{RQ1: What are the different entities and notable practices in the data repair ecosystem?}\\
\textit{RQ2: \newedits{What challenges arise in data repair work? What are the most frequent and unusual challenges, how are they addressed in practice?}}\\
\textit{RQ3: What strategies do data repairers use for their own survival and data repair markets' sustainability?}\\
\textit{RQ4: What implications may the insights from data repair practices have for HCI to mitigate the data poverty concerns of Bangladeshi people and similar other resource-constrained communities?}\\
\end{quote}

\newedits{Our findings reveal that the data repair market has grown and sustains within the computer and related machinery repair market.} In addition to a lack of high-precision data repair machinery and expensive software, the data repair work in this ecosystem is challenged with cross-language learning materials, as well as local knowledge documentation, curation, and propagation. We noted that the repairers restrictively curate the most frequent yet critical repair jobs and the solutions for their future use and for training their disciples. Beyond the well-known problems, there were cases that may not be solved using their curated knowledge or collective peer efforts, and hence led them to turn to other sources, such as foreign forums and LLMs, which enriched our research with insights into repairers' frustration and ethical challenges experienced by them. We also noted the tensions with their survival and the market's growth, which concern pricing, customer relations, and trust among peers.

Our work makes four key contributions. \textbf{First}, we present a detailed ethnographic account of data repair practices in Dhaka, Bangladesh, showing how they operate through human expertise, specialized software, and the socio-political dynamics that shape the field. \textbf{Second}, we map how repair knowledge is produced, propagated, and deliberately restricted within local networks, highlighting practices of secrecy, informal learning, and controlled access. \textbf{Third}, we demonstrate how ``data-poor" societies in the Global South attempt to retain, manage, and restore their data, demonstrating the limits and politics in it. \textbf{Fourth}, we discuss how the practices of the data repair ecosystem, as found in Dhaka, Bangladesh, represent a crucial challenge in the global effort to promote data equity.

\section{Literature Review}
This paper defines \textit{data repair} as the constellation of practices that take into account not only conventional data recovery, such as file recovery from broken phones and damaged memory cards, or fixing cracked screens, but also the diverse forms of access restoration and media conversion of files. Additionally, our concept of Data repair includes regaining entry to accounts through password resets, recovering hacked profiles, and restoring full control, as well as fixing or enhancing digital artifacts by repairing corrupted files, improving picture resolution, converting multimedia formats into usable forms, and even digitally retrieving images and documents from damaged paper copies. By taking these technical, social, and experiential dimensions into account, our definition emphasizes that data repair is not limited to technical recovery but encompasses the broader socio-material work for making data usable again. 

\subsection{HCI and Repair}
Repair has long been a subject of inquiry in HCI and science and technology studies (STS), though often overshadowed by studies of design and adoption. Suchman's analysis of photocopier repair \cite{suchman1987plans, suchman2007human} demonstrated how breakdowns disrupted formalized plans, demanding situated action and improvisation. Orr's ethnography of Xerox technicians showed how repair knowledge circulated through ``war stories" and informal heuristics \cite{orr1990talking}. These insights informed subsequent HCI scholarship that recognized breakdown not as failure but as an ordinary, generative moment in sociotechnical systems \cite{jackson201411}. Jackson and Kang highlighted breakdown, obsolescence, and reuse as central to technological life \cite{jackson2014breakdown}, while Rosner and colleagues examined how repair practices intersect with gender, materiality, and identity \cite{rosner2011antiquarian, rosner2014designing, rosner2014making}. Houston et al. emphasized that repair embodies values of care, responsibility, and sustainability \cite{houston2016values, houston2016caring}, extending its ethical resonance in HCI.

HCI for development (HCI4D) scholarship situates repair within environments of scarcity, where access to new devices is limited and infrastructures are fragile. Jackson et al. described repair ``worlds" in Namibia \cite{jackson2011things,jackson2012repair}, Houston documented inventive infrastructures in Uganda \cite{houston2013inventive}, Ahmed et al. studied mobile phone repair in Dhaka \cite{ahmed2015learning}, and Wyche examined informal information and communication technology (ICT) repair in Nairobi slums \cite{wyche2015exploring}. Collectively, these works show that repair is collaborative, situated, and infrastructural, sustaining access where replacement is unfeasible. Yet most research has focused on hardware and infrastructural breakdowns. Expanding this attention to data repair highlights how informational fragility and recovery form equally critical sites for HCI engagement with resilience and digital justice. This motivates our research to examine how data repair practices extend and complicate existing theories of breakdown and maintenance in HCI.

\subsection{Informality, Learning, and Communities of Practice}
Repair is often enabled through informality in spaces, training, and economies. Polanyi's notion of tacit knowledge \cite{polanyi2009tacit}, Sennett's reflections on craft \cite{sennett2008craftsman}, and McCullough's study of digital craft \cite{mccullough1998abstracting} highlight how skill is embodied, experiential, and often non-codifiable. Suchman and Orr underscored the gap between plans and situated improvisations, and the role of stories in transmitting repair know-how \cite{suchman1987plans, orr1990talking}. Lave and Wenger's concept of ``communities of practice" \cite{lave1991situated, lave2019apprenticeship} emphasized how learning and identity formation are bound together in professional and craft traditions. These perspectives frame repair not as solitary technical work but as collective knowledge cultivated in practice.

Research in repair and access in the Global South illustrates how \textbf{informality functions as infrastructure}. Rangaswamy and Nair examined entrepreneurial repair in Indian slums \cite{rangaswamy2012pc}, Burrell showed how Internet cafés in West Africa became sites of improvisation and critique \cite{burrell2008problematic}, Simone described ``people as infrastructure" in African cities \cite{simone2004people}, and de Bruijn et al. traced mobile access in Africa as supported by ad hoc practices \cite{de2014connecting}. Informality also structures access to tools. Sundaram's Pirate Modernity \cite{sundaram2009pirate}, Karaganis' analysis of global media piracy \cite{karaganis2011media}, and Liang's theorization of the pirate figure \cite{liang2010beyond} highlight how cracked software and pirated infrastructures proliferate in contexts of exclusion. Research on resilience also highlights how communities improvise infrastructures and knowledge-sharing in disrupted contexts, reinforcing the role of informality as a collective capacity to adapt, sustain, and reorganize practices under conditions of adversity \cite{norris2008community, mark2008resilience, mark2009resilience, soden2014resilience}. Together, this literature reveals informality as a system of survival and knowledge transmission that both sustains and constrains repair economies, an observation that motivates closer attention to how digital data repair is organized and reproduced. Being motivated by this scholarship, our research further investigates how informal infrastructures shape learning, knowledge circulation, and repair work in urban marketplaces in the Global South.

\subsection{Privacy, Ethics, and Postcolonial Computing}
Repair also raises acute privacy and ethical concerns. Nissenbaum's framework of contextual integrity \cite{nissenbaum2004privacy} and usable privacy studies \cite{sadeh2009understanding} highlight how disclosure can cause embarrassment, regret, or harm. Yet, unlike medicine (HIPAA) \cite{annas2003hipaa} or law (attorney–client privilege) \cite{raleigh1988attorney}, repair has no institutional protections. Research in ICTD has documented these vulnerabilities. Ahmed et al. showed how repairers in Dhaka handled sensitive personal records \cite{ahmed2015learning}, while Houston described Ugandan repairers confronted with private content \cite{houston2013inventive}. Users often rely on informal trust and secrecy rather than enforceable protections, leaving risks unresolved. Critical perspectives in postcolonial computing emphasize how power and exclusion structure such dynamics. Irani et al. argue that Western-centered models of technology marginalize practices in the Global South \cite{irani2010postcolonial}. Sundaram \cite{sundaram2009pirate}, Liang \cite{liang2010beyond}, and Simone \cite{simone2006pirate} describe piracy as an infrastructure of access and resistance, even as it complicates legality and ethics. These works suggest that repair practices cannot be understood apart from global asymmetries of property, labor, and trust. Situating data repair within these debates illuminates how privacy and piracy interweave in ways that shape not only individual experiences of vulnerability but also collective strategies for technological survival. This motivates our research to ask how repairers negotiate ethical tensions around privacy, trust, and access in postcolonial digital economies.

\subsection{Repair, Values, and Sustainability}
Repair is also a domain where values and sustainability are enacted. Friedman and Nissenbaum's values in design scholarship \cite{friedman1990societal, friedman1996bias, friedman1997software, friedman2007human, friedman2013value} established that technologies embed moral and political consequences. Houston et al.'s Values in Repair \cite{houston2016values} extended this by showing repair as an ethical practice of care and responsibility. Sustainable HCI has emphasized repair, reuse, and repurposing as counterpoints to consumerist discard logics \cite{blevis2007sustainable, mankoff2007environmental, disalvo2010mapping, dourish2010hci}, while collapse informatics reframes breakdown as an opportunity for resilience \cite{tomlinson2012collapse,starbird2013working}. Lepawsky's analysis of e-waste reuse \cite{lepawsky2011beginnings, lepawsky2011making}, Rifat et al.'s ethnography of Dhaka's bhangari recyclers \cite{rifat2019breaking}, and Kim et al.'s reuse framework for obsolete electronics \cite{kim2011practices} further underscore repair's role in sustainability and justice. Studies of digital durability and phone lifecycles extend this perspective, from Odom’s work on sustaining personal digital artifacts \cite{odom2009understanding} to Huang’s investigations of mobile phone replacement, disposal, and sustainability practices \cite{huang2008breaking,huang2009understanding}. Graham and Thrift's theorization of maintenance \cite{graham2007out} further underscores repair's role in sustainability and justice. These insights suggest that repair is not merely technical recovery but a site where value is negotiated, whether through the pricing of services, the prioritization of emotionally significant data, or the cultivation of long-term trust. Practices of secrecy and selective training can also be read as strategies of market survival, linking sustainability to livelihoods as much as to ecological goals. In this sense, data repair highlights how values of care, responsibility, and sustainability are materially embedded in practices, motivating our investigation of how fragile repair economies sustain both data and communities. This motivates our research questions about how values and sustainability are materially articulated through the work of repairing digital data.

\section{Method}
Our six-month-long ethnography took place during Dec 2024 - Jan 2025 and May 2025 - Aug 2025 in Dhaka. We engaged with two types of data repair places: eleven low-fidelity neighborhood repair shops and ten high-fidelity lab-work-based repair outlets in bigger marketplaces. The neighborhood shop repairers restore physically damaged photographs and official documents, unlock phones, and help with social media account recovery and OTP issues, while the lab experts at the high-fidelity labs specialize in professional hardware and software repair. We also engaged with mediators who handle customer interactions and transfer devices from customers to the data repair lab experts. The low-fidelity repair shops were in Mirpur, Dhanmondi, Badda, Kathalbagan, and Gulistan neighborhoods. The high-fidelity data repair outlets were in six marketplaces: Multiplan Center, Suvastu Arcade (ICT Bhaban), IDB Bhaban Mirpur, Motalib Plaza, Gulistan Repair Market, and Eastern Plus Shopping Complex.

\subsection{Participant Recruitment}

We recruited participants by distributing flyers and directly reaching out to them. We explained the research and sought opportunities to observe their work and interview them. We recruited neighborhood shop repairers by approaching them directly in their shops. The initial visits to the markets helped us communicate with the mediators, who often act as the first point of contact for customers seeking repair and then carry the devices to specialized labs.  These mediators also participated in the study and introduced us to repair lab experts, facilitating access to otherwise closed workspaces. Also, often participants referred others who might be a good fit for the study \cite{goodman1961snowball}. All the participants were 18 years or older with at least six months of prior experience in data repair (e.g., data repair lab expert, mediators, neighborhood repair shopkeeper). See Table~\ref{Tab:demo} for participants' demographics. We did not find any female repairers. This is because the places where repairers work (see Fig.\ref{fig:market}) are seemingly culturally inappropriate and unsafe for women to work, and the profession itself is not considered lucrative for women, as bystanders opined during our observations. See more noted in subsection \ref{subsec:data-repair-people}.

\subsection{Observation}
Our observation sessions took place in eleven low-fidelity neighborhood repair shops and ten high-fidelity lab-work-based repair outlets in six marketplaces. For the sessions with the neighborhood repair shops, we scheduled the sessions in advance and observed them work during their work hours. For high-fidelity repair observation, we first visited and observed the mediators' desks in the market and then the specialized labs of data repair experts. The shop owners and mediators facilitated our access to these spaces upon our request. Depending on the case and required expertise, a mediator coordinates with one or more lab experts. They scheduled convenient times for us and introduced us to data repair lab experts. Upon our explaining the goal of this work, the participants asked for clarification before they agreed to be observed. We conducted 16 three-to-four-hour-long observation sessions (almost 50 hrs in total) to understand their repair processes, the tools and software they used, and the ways they interacted with customers. We took detailed notes in our notebook during the sessions and, with participants’ permission, took photos of their workspaces and repair setups.


\begin{table}[t]
\centering
\small 
\setlength{\tabcolsep}{4pt}
\begin{tabular}{|rl|}
\hline
Total Participants: & 28 (Female: 0, Male: 28)\\
\hdashline
\multicolumn{2}{|c|}{\textbf{Participant Roles}} \\
\hdashline
Data Repair Lab Expert: & 9 (Female: 0, Male: 9)\\
Mediators: & 8 (Female: 0, Male: 8)\\
Neighborhood Repairers: & 11 (Female: 0, Male: 11)\\
\hdashline
\multicolumn{2}{|c|}{\textbf{Age Range (in Years)}} \\
\hdashline
All: & 19--45, median 28\\
Male: & 19--45, median 28\\
Female: & N/A , median N/A\\
\hdashline
\multicolumn{2}{|c|}{\textbf{Years of Experience in Data Recovery}} \\
\hdashline
Data Repair Lab Expert: & 2--11 years, median 7\\
Mediators: & 1--8 years, median 4\\
Neighborhood Repairers: & 1--8 years, median 4\\
\hdashline
\multicolumn{2}{|c|}{\textbf{Education}} \\ 
\hdashline
Primary School: & 6 (Female: 0, Male: 6)\\
High School: & 4 (Female: 0, Male: 4)\\
Secondary School Certificate: & 5 (Female: 0, Male: 5)\\
Higher Secondary Certificate: & 7 (Female: 0, Male: 7)\\
Technical Diploma: & 5 (Female: 0, Male: 5)\\
Bachelor’s Degree: & 1 (Female: 0, Male: 1)\\
\hline
\end{tabular}
\caption{The demographic details of the participants}
\label{Tab:demo}
\end{table}

\subsection{Semi-structured Interviews}
We conducted 28 semi-structured interviews \cite{ruslin2022semi}. All interviews were conducted in Bengali, the native language of both the participants and the researchers. At the beginning of each interview, we explained the purpose of the study and sought informed consent. During the interviews, we asked participants about their entry into data repair and the social conditions that shaped these trajectories, the ways they built and shared skills, their experiences of challenges and the tactics they developed to navigate or resist these challenges. Each session lasted between 45 to 60 minutes and was scheduled according to the participant’s availability. All Interviews were held in person at the repair venues. We took detailed notes. Twenty-three participants gave us permission to audio-record their sessions. These materials were later transcribed and translated for open coding and thematic analysis \cite{fereday2006demonstrating, boyatzis1998transforming}.

\subsection{Data Collection and Analysis}
We collected around 150 pages of fieldnotes, more than 200 images, and 18 hours of audio recordings. We transcribed the audio recordings and translated them from Bengali to English, and removed identifiers before conducting an open coding and thematic analysis \cite{fereday2006demonstrating, boyatzis1998transforming}. Two members of the research team independently reviewed the transcripts to become familiar with the content. During the open coding process, we allowed codes to emerge inductively, capturing participants’ reflections, stories, and critiques of their work and the repair ecosystem. The initial codes included topics such as software barriers, undocumented repair challenges, and community knowledge sharing. \newedits{We then clustered related codes into higher-level themes based on conceptual similarity and how these patterns repeatedly appeared together in participants’ narratives. These themes emerged naturally from the data through multiple rounds of comparison and discussion within the research team} (see codebook in supplementary materials). These themes are presented in the next section.

\subsection{Ethical Concerns and Positionality}
This study was reviewed and approved by our university's Institutional Review Board (IRB). All participants provided informed consent and were assured confidentiality. As researchers, our positionality shaped the study: all the authors are Bangladeshi and shared linguistic and cultural familiarity with participants, which helped establish trust. We acknowledge that our academic and socioeconomic positions differ from those of participants, and we sought to mitigate these imbalances through long-term engagement and reflexive analysis.

All authors are Bengali and Bangladeshi, born and raised in Bangladesh, with deep familiarity with the cultural, linguistic, and social contexts of repair work. Our long-term engagement with community-centric and infrastructure research grounds this study, with multiple authors bringing over a decade of ethnographic experience. This shared background enabled us to build trust with participants, while we remained reflexive about the differences between our academic positions and the participants' lived realities.

\section{Data Repair Ecosystem in Dhaka}
The data repair ecosystem in Dhaka includes both low-fidelity repair shops in neighborhoods and high-fidelity lab-work-based repair outlets in bigger marketplaces and people associated with them. This section details the venues, the people, the equipment used, and the service offered. 


\subsection{Data Repair Venues}
\begin{figure}[!t] 
  \centering
  \includegraphics[width=\linewidth]{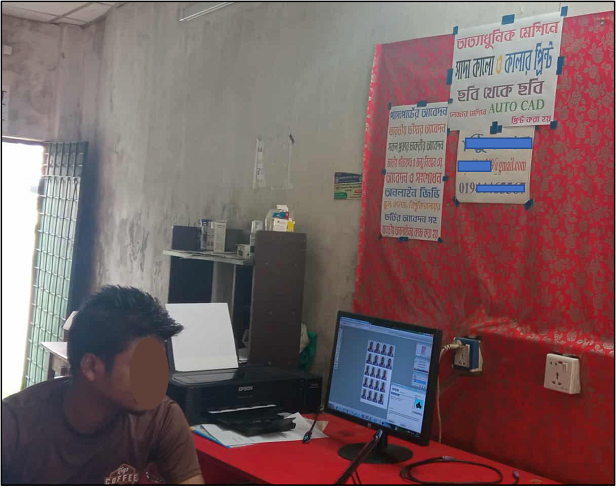} 
  \vspace{-15pt}
  \caption{Workstation of a typical low-fidelity neighborhood repair shop. The repairer was sitting before a computer screen, the wall behind the screen hung his shop's advertisement flyers with his contact information, describing the services he offers, including retrieving damaged photos. The photo was taken when the repairer was retrieving a passport photo from a damaged paper copy. The damaged photo is of the customer's mother, who has passed away, but the customer still needed it for paperwork.}
  \label{fig:nrs}
\vspace{-10pt}
\end{figure}

There are an uncountable number of \textbf{low-fidelity repair shops} in the Dhaka neighborhoods, possibly one every half square mile. The most common repair requisitions that come to them include fixing the resolution, improving clarity and visibility, and restoring the quality of old, torn, water-damaged, or faded physical copies of photos and official documents like damaged certificates, land deeds, etc. (see Fig.\ref{fig:nrs}). They also help customers unlock mobile phones or social media accounts when passwords are forgotten, and solve one-time password (OTP) problems. Complex tasks such as retrieving damaged, corrupted, or deleted files from broken or damaged phones, memory cards, tablets, or hard drives, or lost files from web spaces are generally beyond their capacity. The neighborhood repairers typically send these customers to high-fidelity repair shops.

\begin{figure}[!t] 
  \centering
  \includegraphics[width=\linewidth]{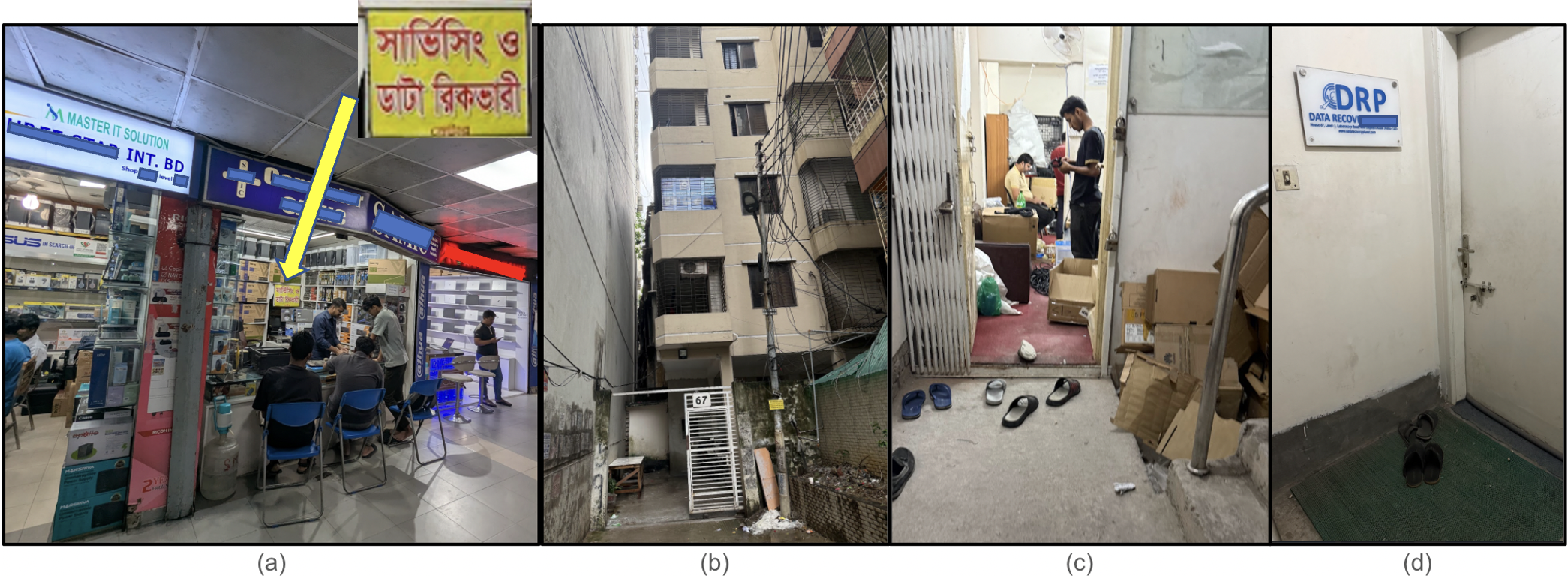} 
  \vspace{-15pt}
  \caption{(a) One of the high-fidelity repair outlets among a cluster of small computer shops inside the Dhaka Multiplan Center. It has a small yellow Bengali signboard reading ``Servicing and Data Recovery." Customers often discover such shops by chance while browsing and approach them for data recovery needs. In this scene, a mediator in a blue shirt is negotiating with customers outside the shop and receiving their devices for recovery. (b) A narrow alley leading to an apartment building where actual data recovery work takes place. On the third floor, a data repair expert rents a flat to perform recovery tasks, located a short distance away from the main markets. (c) A small, crowded apartment-based lab where a lab expert and his team handle devices and tools related to data repair. (d) The entrance of another lab where data recovery takes place, with shoes left outside as they are not allowed inside.}
  \label{fig:market}
\vspace{-10pt}
\end{figure}

The \textbf{high-fidelity repair outlets} have two distributed work stations: customer attending shops and repair labs. Among many computer or phone sales and repair shops, only a few offer a corner where one or two salespeople sit and attend to data repair, seeking customers. Thus, the \textbf{customers attending shops} are located inside crowded large computer or phone markets and are more visible. The shops that offer a corner for data repairers' attendees (mediator, henceforth) would place a signboard or poster that says they also provide data repair and recovery support (see Fig.\ref{fig:market}(a)). The mediators talk with customers, check the device, negotiate the price, and record details like phone number and the customers' National Identity (NID) card. These shops receive the repair work orders and also hand over the devices upon fixing. The easiest way to find a data repair point is to ask around in a computer and mobile phone marketplace, and the shopkeepers would usually point you to one of these specialized places. These shops' open hours follow the markets' schedules: usually open 10 am - 8 pm. 

On the other hand, the \textbf{repair labs}, where the actual tasks of repairing take place, are rare to find as they are not located in the marketplaces. We visited six repair labs, which were within 10-20 minutes walking distance from the main marketplaces, where their mediators were. The batch of repair-seeking devices is delivered to the lab every evening by the mediators. All the labs were set up in rented apartments or small buildings (see Fig.\ref{fig:market}(b)). These labs have computers, repair tools, and software required for data repair. Given that the tools are expensive, highly sophisticated, and delicate, the labs were very well maintained. The rooms are air-conditioned, and windows are never opened to prevent dust. No shoes are allowed inside the lab (see Fig.\ref{fig:market}(d)). Customers' or other people's entry is also restricted. Our repetitive visits for the ethnography helped us build rapport, and that is how we were only allowed inside the labs for brief observations.


\subsection{Data Repair People} \label{subsec:data-repair-people}
\textbf{Low-fidelity repairers} with neighborhood shops are the most common figures in Dhaka's repair landscape. None of the shop repairers we interviewed had formal technical training, and at most five of them had completed a Higher Secondary School Certificate. Seven of them began their career by selling phone accessories, offering minor mobile servicing, or running photocopy and printing stalls. For example, P14 explained that he started by printing ID photos and documents in his small shop, but customers often asked him to restore damaged photos as well. He began experimenting with basic editing software and gradually built up his skills in this area. Similarly, P7 used to work in a mobile phone servicing shop, and now he offers phone servicing along with unlocking phones with forgotten passwords in his new shop. These varied backgrounds provided them with an entry point into data repair work.

The high-fidelity data repair work is distributed between two groups: data repair lab experts and mediators. The \textbf{data repair lab experts} are technicians who do the actual process of retrieving information from damaged or inaccessible devices. Four of our lab experts said they had come into this field with some technical background. For example, participant P9 explained that he had his own IT business before joining this work. He used to set up hardware like Closed-Circuit Television (CCTV) systems, Private Automatic Branch Exchange (PABX) lines, and other IT services. His first encounter with data recovery came when a customer asked if he could also recover data from hard drives. At first, he did not know much about it, but as more people asked, his curiosity grew, and he started researching tools. This curiosity slowly brought him into the recovery field. Three experts entered the field because of personal experiences. Participant P1 shared that in 2008 his father’s office computer crashed, and five years of archive data were lost. He remembered feeling helpless because no one in Bangladesh could provide such a service at that time. That frustration stayed with him, and it motivated him to learn recovery so others would not face the same problem. Despite such different pathways, the field is almost entirely male-dominated. In our study, we did not meet a single woman repairer. This absence reflects broader gender gaps in Bangladesh, where repair skills are seen as men’s work, markets are socially unwelcoming for women, and families encourage women to pursue more formal professions. For those who do join, training is a necessary step. Most experts learn from official vendors such as SEI, MRT, or DFL labs, who sell both the recovery tools and paid training. The training is expensive, and some senior technicians also offer informal training, though at very high fees. For example, participant P17 said it could cost up to BDT. 500,000 (about USD. 4000) to learn directly from him.  

The \textbf{mediators}, on the other hand, are the people who sit at the customer-attending shops in crowded marketplaces. They talk with customers, check the device, negotiate the minimum price for recovery, and record details such as phone numbers or NID cards. The mediators act as the bridge between the visible shop and the hidden lab. All the mediators we met had previously run regular shops inside the shopping complex for several years, making them well-known figures in those markets. Data repair lab experts collaborate with these mediators through fixed agreements.



\subsection{Data Repair Community} \label{subsec:data-repair-community}
All nine lab experts told us about their \textbf{community chat groups on Telegram channels}. These groups are geography-insensitive and more task-focused. Hence, repairers from not only Dhaka, but across Bangladesh come and join the community over the Telegram channels. They post to seek solutions to the repair problems they think might be common, but they struggled to solve; ask and enlighten about new software and hardware; share online resources they came across and thought might be useful to other colleagues; and brand themselves with their expertise, exclusive even within the repair community. This self-branding and product advertisement often helps other repairers to find repairers with context-specific skills and appropriate tools and techniques to find solutions to the problems. Several repairers told us that conversations on these channels have led to innovative solutions many times, which have pioneered the data repair community in Bangladesh. 


\begin{figure}[!t] 
  \centering
  \includegraphics[width=\linewidth]{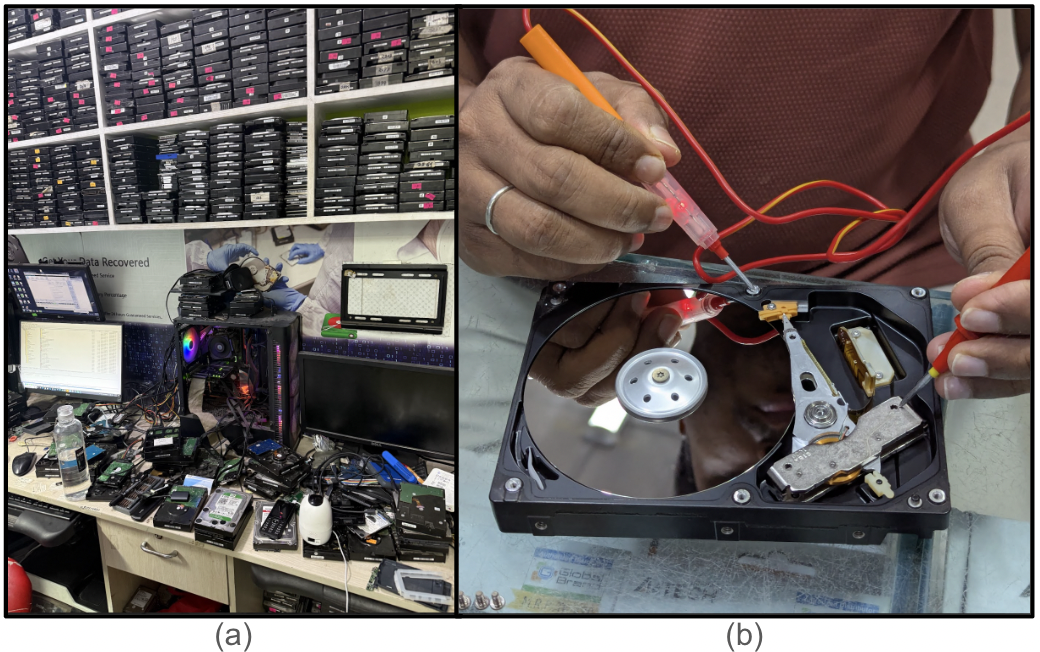} 
  \vspace{-15pt}
  \caption{(a) Inside a data recovery lab, shelves are filled with hundreds of labeled hard drives that serve as donor parts for recovered data. On the workbench, multiple patient drives are connected to a recovery workstation with adapters, power leads, tools, and a USB microscope. On the left, two computers are running: the top screen is used for diagnostics and cloning, while the bottom screen displays recovered files for verification and transfer. (b) The expert is testing the current flow on the opened hard drive's connection points with a test lamp. The glowing red light shows that the circuit has continuity, helping him identify the faults.}
  \label{fig:lab}
\vspace{-10pt}
\end{figure}

These groups are protected, invite-only, and do not have any open join or invite link for self-invitation. They are added to the community through peer connection and introduction. While they did not give us the exact name, they referred to those with a nearby pseudonym, and we figured there were at least four such Telegram communities for them to grow peer-based knowledge.

\subsection{Software and Hardware}
Low-fidelity repairers use a variety of tools in their neighborhood shops. For repairing physically damaged photos, they use scanners for digitization, Adobe Photoshop \cite{adobe-photoshop} for core editing, and ScanTailor Advanced \cite{scantailor-advanced} for cleaning up documents. Recently, some repairers have also started using Topaz Photo AI \cite{topaz-photo-ai} in their recovery work. For unlocking devices, the most common tools mentioned are Octoplus \cite{octoplusbox} and ChimeraTool \cite{chimeratool}. In cases of social media password recovery or OTP-related issues, they generally follow the official guidelines provided by Facebook or Instagram, which do not require any specialized software. Repairers obtain the necessary hardware from local vendors and purchase software subscriptions from official websites for their work.

The repair lab expert requires expertise in both software and hardware. The \textbf{software} involves a range of specialized tools that allow technicians to diagnose and recover data. Commonly mentioned software included PC-3000, Victoria, Disk Drill, EaseUS, DMDE, and R-Studio \cite{rstudio-datarecovery, pc3000-acelab, victoria-hddssd, diskdrill-cleverfiles, easeus-drw, dmde}. Access to these tools varies. Some repairers purchase them from official vendors, but many rely on pirated versions shared through closed Telegram groups known only to recovery personnel. The pirated versions are cheaper but unstable, often breaking after updates. Learning these tools is not straightforward. Formal vendor courses may last for weeks, yet real proficiency develops only after long practice. As P16 noted, the time required to complete a recovery is case-dependent: simple issues may take hours, but complex or severely damaged devices may require days or even weeks. Also, the repairers' skills play a role. 

The \textbf{hardware} aspect is equally critical because recovery often requires physical interventions alongside software. Repairers typically keep desktop computers, laptops, donor hard drives, soldering kits, and chip programmers in their labs. These items are commonly available in local markets, and technicians use them to perform delicate tasks such as soldering circuits, replacing damaged hard drive heads, or fixing broken connectors. Along with these standard tools, repairers also rely on more specialized hardware such as Spider boards \cite{acelab-spider-board-teeltech}, which allow direct data access from chips. When purchased through official vendors, these boards are often bundled with licensed software, but many technicians turn to informal back channels to import them at lower cost. These routes provide affordability but also carry risks of unstable supply, counterfeit parts, or lack of after-sales support. Since these repair practices overlap with common phone and computer servicing, many technicians build on their earlier technical experience or learn directly from nearby repair shops. This makes hardware skills more embedded in the everyday repair culture of the marketplaces. By combining these physical interventions with software-based tools, technicians expand their capacity to handle cases that software alone cannot solve.

\subsection{Services vs. Price}
The \textbf{pricing practices} follow a tiered structure that reflects the complexity of the case. In neighborhood shops, the cost of recovering damaged photos and documents depends on the physical condition of the file and typically ranges from BDT. 200 (USD. 2) to BDT. 1000 (USD. 8) per photo or per page of a document. Unlocking mobile phones typically costs around BDT. 3500 (USD. 30). For recovering social media accounts or OTP-related issues, there is no fixed price. These cases usually involve low-tech-skilled customers, and repairers often provide the service for a very minimal fee (around USD. 1) or sometimes free, since they require little additional effort or time.

High-fidelity repair outlet situations are a bit different. Mediators usually give customers a minimum starting price, but they explain that the final cost depends on the condition of the device, which cannot be fully known at first glance. For simple cases such as accidental deletion, formatting, or corrupted file systems, the minimum price starts at about BDT. 3500 (USD. 30). If the drive is working but has electronic problems such as bad sectors or minor Printed Circuit Board (PCB) issues, recovery usually starts at around BDT. 7000 (USD. 60). Recovering data from pen drives or small memory cards costs about BDT. 6000 (USD. 50), while video recovery from Digital Single-Lens Reflex (DSLR), GoPro, drone, or RED camera cards begins at BDT. 10000 (USD. 80). The most expensive cases involve Solid-State Drive (SSDs) or devices with both hardware and logical corruption. All of these cases represent only the minimum prices quoted by mediators to customers. Once a device reaches an expert lab, the specialist assesses its actual condition and then calls the customer to confirm the final cost of recovery. At this point, the customer decides whether to proceed. If they agree, the expert begins the recovery process; if not, the expert returns the device to the mediator, who hands it back to the customer unrecovered.

During our ethnography, we observed that mediators do not take a device from a customer until the customer confirms they have the \textbf{minimum budget} required. The actual recovery cost may be higher than this amount, but the minimum is a condition for acceptance. Mediators explain that they will only keep the device and proceed with screening if the customer agrees to the minimum; otherwise, the customer has to leave with the device. 


\section{Findings}
Our thematic analysis revealed different professional strategies that the data repairers adopt to solve the challenges they encounter during repair, as well as to sustain in the market to date and in the long run. We categorize and explain them below under four sub-themes: addressing known and documented challenges with software and tools, addressing novel and exotic challenges with software and tools, complying with market demands and valuating the value of data, and collective protocols for sustaining the data repair market.  

\subsection{Addressing Known and Documented Challenges with Software and Tools} 
Some of the data repair requisitions that repairers address are already well known in the community, as they frequently appear in repair work. Over time, repairers have investigated these recurring problems, developed and documented reliable workarounds, and circulated them among peers. This way of documentation contributes to community-centric knowledge curation in data repair.

\subsubsection{Challenges with Locally Manufactured Phones}
Seven lab experts mentioned that many traditional and internationally renowned repair software fail to recover data from locally manufactured phones. Local companies like Walton and Symphony produce affordable phones that attract a large number of local low-resource and low-income users. While these phones are popular, their internal structures are incompatible with standard data repair tools, creating significant challenges for repairers. Lab expert P9 gave us examples that phone models such as the Walton Primo X5 use MediaTek chipsets and security settings that block repairers' attempts to gain low-level access. Unlike most other phones that expose the internal storage as a disk, Walton Primo X5 phones use Media Transfer Protocol (MTP) for transferring files, which prevents most standard recovery tools from functioning and requires additional advanced specialist steps for repair. Continuing from his point, his colleague P10 in the same lab explained to us why this mismatch happens between data repair tools and devices.

\begin{quote}
\textit{``To make phones affordable, companies manufacture, produce, and refurbish them locally. Since they need to be cheap, many components are either missing, reduced in number, or of low quality. That is why data recovery tools fail to recover data from these devices properly. Also, when we work on these phones, we have to be extremely careful so that no accident happens and the recovery process can be done properly.'' (P10)}
\end{quote}

This response shows how the mismatch between locally manufactured phones and software causes a significant challenge for data repair. When we asked how lab experts deal with these issues, they explained that the solutions are already well known among their peers in most cases. Because these cases are so common, other repairers have likely faced them before and figured out workarounds. P1 shared how he handles these cases through community documentation.

\begin{quote}
\textit{``These issues with cheap phones are very common, and we all know that. When working with them, we use service mode or BootROM workflows to load a suitable agent and read the data. If that is not possible, we do a chip-off read of the memory and rebuild the data from the phone. However, success is not guaranteed, as we fail more than 50 percent of the time due to low-quality components.''(P1)}
\end{quote}


Thus, insights and experiences with common task-level challenges are turned into shared knowledge and are often shared on \textbf{community chat groups on Telegram channels} (see more in subsection \ref{subsec:data-repair-community}). Even though unsupported phones create repeated obstacles, repairers have learned to manage them through documented solutions and community practices. In this way, the community supports its peers to keep up the pace and sustain. 

\subsubsection{Compatibility Challenges with Old Devices and Software}
Six lab experts said they often face challenges when working with outdated hard drives, memory cards, and other storage devices. Most low-income users in Bangladesh often avoid spending extra money on new devices as long as the ones passed down to them by a family or friend are still functional. However, data repair work is challenged when they work with such phones if they are of outdated models. The repairers told us that old devices are often compatible with older versions of repair software like PC INSPECTOR Smart Recovery v4.5, Lexar Image Rescue 3 etc. Different generations of repair software are compatible with hardware from those and the nearby generations. But these older software versions also create problems because they were originally designed to run on old computer operating systems (OS) like Windows XP or Windows 7, which were popular in the 2000s. When repairers try to run them on newer versions of Windows (Windows 10 and advanced), the tools often fail due to mismatched drivers or incompatibility issues. P16 explained this to us with an example,

\begin{quote}
\textit{``I remember once I got an old 5 V SmartMedia card (an early higher-voltage version of a solid-state flash memory card, which was popular in the 2000s) for recovery, and the recovery tool did not work. Probably because the card's reader or driver had compatibility issues with Windows. When I shared this problem in the group, they suggested that since the card was such an old model, I should try older software like PC Inspector (widely used during the early to mid-2000s) on Windows 7 (a popular computer OS in the early to mid-2010s). When I tried it with that tool, it worked perfectly.'' (P16)}
\end{quote}

This example shows how repairers face repeated obstacles when working with outdated devices. To deal with these cases, repairers have developed a strategy of keeping old computer OS and older versions of software stored and installed. As lab expert P10 described:

\begin{quote}
\textit{``Sometimes we get outdated models of devices for recovery, which is why I intentionally keep one of my computers running on Windows 7. For old devices, we need old tools, so we make sure to keep those tools stored so that if any one of us needs the software, they can download and install it to carry out data recovery from outdated devices.'' (P10)}
\end{quote}

All of the repair lab experts told us about this strategy. They shared the risk of getting these OSs and software accidentally updated via auto-update from the internet, which amplifies mismatching troubles. Five lab experts also mentioned that they face significant challenges in running these outdated operating systems on their machines. As P4 explained:

\begin{quote}
\textit{``Drivers, browsers, USB firmware, and networking protocols keep getting updated, and newer versions often do not work on Windows 7, so we block those updates. Using the internet is also difficult because browsers do not run properly, many sites refuse to open because of missing security support, and there is a constant risk of viruses. For this reason, we usually try to avoid using the internet on these machines and keep the  computers completely offline.'' (P4)}
\end{quote}

The participants informed us that these updates also cause trouble with cracked versions of software. They added that they intentionally kept the Auto-Update features of the old OSs, and that software was turned off to avoid unintended troubles. All the data repairers said they had specifically documented the step-by-step solutions associated with old versions of operating systems and software. Additionally, they also noted that maintaining an extra computer in their lab for recovering data from older devices was a financial and space burden. 


\subsection{Addressing Novel and Exotic Challenges with Software and Tools} 
The repairers often face novel and exotic repair problems that are not documented on peer platforms, and solutions are unknown. Many times, these efforts create new difficulties. This theme shows how repairers deal with undocumented challenges and the strategies they use when they have no support from their community.

\subsubsection{Challenges with Hardware Encryption and Proprietary File Formats}
Sometimes a problem is not documented in the community, and lab experts do not have a clear way to move forward. These cases are new or unusual, and even consulting peers offers no solution. In such situations, repairers try different workarounds, but many times they still cannot solve the problem and have to return the device. These cases show the limits of shared knowledge and the points where the community does not yet have a solution to offer. As P18 said,

\begin{quote}
 \textit{``Many times we receive devices where all data is encrypted in hardware on the USB bridge board (an electronic circuit board that acts as a translator, allowing a USB-compatible device to communicate with a component or device that uses a different data protocol) by default. If the USB–SATA bridge(a converter device that allows a hard disk drive (HDD), solid-state drive (SSD), or CDs, DVDs, and Blu-ray discs to connect to and communicate with devices that have USB ports)  burns out and you remove the bare SATA drive, the contents only appear as ciphertext. In our community, no one has developed a method to recover data for such cases. The only option is to return the device to the owner, as there is no other solution.'' (P18)}
\end{quote}

In non-geek terms, the problem was that the external hard drive has a special secret built-in feature that encrypts all data as it is saved. This encryption is managed by a tiny chip on the board inside the drive's case, i.e., the physical jacket of the device. With a USB bridge board, the decryption key is lost forever. Hence, the expert data repairer could not help because there is no way to get the combination back. Even when repairers can image the drive, they cannot make the data readable without the original bridge or keys. With no community workaround to follow, the case ends without success. Five lab experts also shared that they face format barriers during data recovery. Their standard tools fail to rebuild the files without the original system or vendor players, and their community has not found a workaround to bypass this issue. As P2 puts it,

\begin{quote}
 \textit{``Sometimes I get CCTV clips that were exported from low-cost DVRs as .264, .h264, or .dav files. These files usually come without the H.264 headers and the DVR’s own index, so when I try them in VLC or FFmpeg, all I see is a black screen. Unless I have the original DVR or the vendor’s player to rebuild those headers, there is no particular way to make the clips playable.'' (P2)}
\end{quote}

Both cases highlight undocumented challenges where the community does not yet have a solution. Repairers try what they can, but encryption tied to hardware and missing proprietary video structures leave them with no path forward. In these situations, they document the failure, inform the client, and return the device. These unsolved cases mark the boundary of current community knowledge and show where further investigation is still needed.

\subsubsection{Challenges with Web Search and Large Language Models} 
When repairers cannot find a solution in the community, they often try seeking help from the web. They try Google first with both Bengali and English keywords, and they know about the problem. But often they can not put the actual technical words that are understandable or searchable by Google in order to give them proper solutions. They also anticipate that other countries that have a repair culture, such as China and Russia, might have resources for them, but often suggestions in Chinese and Russian languages do not work because of \textbf{language barriers}. Upon preaching the possibility of Large Language Models like ChatGPT and Gemini being helpful in this regard, the participants shared their diverse experiences with them. Seven lab experts told us that they had tried ChatGPT (they termed all LLMs as ChatGPT; for example, Gemini is referred to as "Google's ChatGPT") at least once when all other options failed. However, most of these attempts led to frustration. Five lab experts explained that ChatGPT often gives wrong or hallucinated responses, which creates even more challenges for them. Because repairers only turn to ChatGPT when no other solution is available, a wrong answer at that stage is very risky. For already damaged data, such errors can worsen the damage, reduce the probability of data repair, and even cause accidents with the device. As P13 said, 

\begin{quote}
\textit{``Once I asked ChatGPT how to recover a corrupted SanDisk SD card that UFI Box was not detecting. It confidently told me to wire through the ISP pinout and enable a ‘virtual drive’ from Device Manager. Those options do not exist in my UFI setup, and I lost half an hour chasing settings. Later, I confirmed with our community that no such function exists. Since then, I only use it for very basic descriptions and avoid asking technically detailed questions.'' (P13)}
\end{quote}


Three lab experts also said that while ChatGPT can sometimes be useful for very basic tasks, it lacks the deep and specific knowledge needed in data repair work. Repairers believe that because they keep their knowledge highly restricted, ChatGPT lacks data repair-related information and gives subpar responses, as P10 explained:

\begin{quote}
\textit{``In data recovery work, we mostly learn by discussing among ourselves, experimenting, and sharing knowledge within our own circle. I think ChatGPT does not have access to this kind of specialized knowledge, and that is why it often lacks enough understanding to give us proper answers in this domain. So, I suggest my peers avoid using ChatGPT for data recovery work.''(P10)}
\end{quote}

Some lab experts faced a different kind of problem when trying to use ChatGPT for undocumented cases. The moment ChatGPT recognizes that the \textbf{software in question is cracked or pirated, it immediately stops answering}. Instead of giving technical guidance, it responds \textbf{with a policy warning}. For repairers who are already stuck because the information is not documented anywhere else, this creates another obstacle and blocks them from getting the help they need. As lab expert P18 described, 

\begin{quote}
\textit{``I was working on a task, and at one point, when I was stuck, I sought help from ChatGPT. I mentioned the software version so that ChatGPT could understand how to solve the problem with that software. As soon as it realized that the software was cracked, it immediately turned off the response and said discussing this was outside their policy or something.'' (P18)}
\end{quote}

All seven repairers shared similar ethical conflicts with ChatGPT or other LLMs, as these tools did not recognize the repairers' needs for using cracked and pirated versions of software, and not only refused to help but also tried to scare them with policy and ethics references. 

These examples highlight how repairers struggle when they turn to LLMs as a last resort. \newedits{These are undocumented challenges where no ready solution is available in the community. When repairers search for workarounds, ChatGPT often provides incorrect guidance or stops responding as soon as it detects cracked or pirated tools. These refusals follow Western intellectual property norms, which do not align with the practical realities of Global South repair work. Instead of helping, this disconnect creates an ethical conflict for repairers and pushes them back to their peers or to their own experiments.}

\subsection{Complying with Market and Valuing the Values of Data} 
For repairers, sustaining in the market is always difficult because the customer base is limited, and customers are also reluctant to spend on data repair. If the repairers set their prices too high, most customers will simply walk away, and the repairers will end up with no work. Because of this, repairers have developed careful strategies to adjust their charges depending on the type of data. These strategies allow them to survive in a market that is still growing. This theme shows how repairers use tactical pricing as a short-term way to keep their work going and sustain themselves in the field.

\subsubsection{Pricing Based on Emotional Value of Data}
Fourteen repairers said that one of their sustaining strategies is to adjust the price depending on the customer's personal financial situation and emotional connection with the data. Neighborhood shop repairer P20 said that people from many different backgrounds come to them for data repair, often with strong emotional connections to the data, such as paper or digital copies of old photos with deceased family members (see a similar example in Fig.\ref{fig:nrs} caption), educational certificates, or legal papers in old formats. Depending on the value of the data, repairers sometimes adjust their prices. P14 explained one such case where he did the repair work for free for one of his customers.

\begin{quote}
\textit{`` Once, a middle-aged rickshaw puller came to me with a torn, faded paper print of his father he wanted restored. The picture was in very bad condition, and he did not have enough money to repair it. He told me that the night before, he had seen his father in a dream and was missing him deeply. That damaged picture was the only memory he had of his father, and if it was lost, he would have nothing left. His words touched me because I also lost my father, so I decided to do the work for free. In cases like this, I do not think about money, because helping someone may later bring me more people through their referrals.'' (P14)}
\end{quote}

This response shows that repairers place strong emotional value on data. Although they sometimes reduce the repair cost, their strategy is that a customer who pays a small amount may later bring more clients to them. However, in some cases, repairers also increase the price when they see that a customer has a good financial background and the data carries strong emotional value. They know that if the data contains memories that cannot be recreated, customers are often willing to pay more. One such example was shared by lab expert P9,

\begin{quote}
\textit{``Once I got a memory card where there were photos and videos of a couple. The mediator told me that they were newly married and they had gone to the sea in Cox’s Bazar for their honeymoon. Their phone accidentally fell into the water, and the memory card inside the phone was damaged. The card contained all their wedding photos, and the pictures they took during their honeymoon were stored on that card. I understood that those pictures had a lot of emotional value, and I knew that no matter how much I charged, they would pay for it. So, in that case, I asked for a bit more money to recover the data because those pictures had emotional value to the customers.'' (P9)}
\end{quote}

Both examples highlight how repairers use the emotional value of data as a short-term strategy to sustain their work. This approach helps them survive and turn difficult market conditions into opportunities by carefully reading the situation and understanding the depth of someone's loss or attachment.

\subsubsection{Strategies Based on Likelihood of Customer Repetitive Visit}
Twelve repairers said they use judgment of the customers' background and tech skills as a strategy to sustain their work. Instead of relying on fixed prices, they try to understand who the customer is, how they use their phones or devices to be repaired, how well they maintain those, how likely that customer is to break the device and come back again, and how much trust can be built with them. By doing this, repairers decide their charges for the repair task, often offering a \textbf{discount for loyalty} to potential repeat customers. Lab expert P16 explained how he uses the educational background of a customer to decide the price-

\begin{quote}
\textit{``I try to judge whether this customer is the type who might lose their data again. Educated customers usually take backups after losing data once. They learn from that experience and know how to store and back up their files, so they are usually one-time customers. I know they probably will not come back, so I intentionally charge them a little more. On the other hand, less educated customers or those who do not understand data storage often make the same mistake again. There is a good chance they will return to me in the future, so I deliberately charge them less to encourage them to come back. They are more likely to become long-term customers.'' (P16)}\end{quote}

Seven lab expert repairers pointed out that most people who come for repairs have low technical skills and limited income. They see this as an opportunity for their business in the market. As P17 said,

\begin{quote}
\textit{``Most people I talk to do not even know what cloud storage is. Even if some do, after the 15 GB free limit, they do not want to buy extra space because the subscription cost is too high for the average person. On top of that, purchasing storage requires a credit card, and most people in our country do not have one. As a result, they keep their data on their devices, and when those devices fail, they come to us for recovery, which becomes our source of income.'' (P17)}
\end{quote}

\subsubsection{Pricing for Corporate Clients}
Seven lab experts reported charging higher fees for cases involving data from organizations or offices. In Bangladesh, customers often bargain heavily for personal data repair tasks; they are ready to compromise quality service and would even let go of repairing if they are not convinced by the offered price. However, corporate clients are different; getting the data repaired is not optional for them, and they do not compromise the quality. Repair work with official data often involves repairing and recovering confidential files with sensitive information, and without which the functionality of the organizations would stop or projects would get delayed. Repairers know that companies are bound to recover their data no matter the cost, which gives them a chance to raise the asking price. As participant P9 explained, these jobs require extra effort and carry higher risks compared to personal data repair:

\begin{quote}
\textit{``Whenever organizations or offices bring me data recovery work, I always keep the charge higher because this type of data is highly sensitive and usually critical for their daily operations. They also require proper documentation, signatures, receipts, and expect every single file to be recovered properly. These jobs come with more risk, strict requirements, and much higher pressure, so I often charge two to three times more than I would for personal data.'' (P9)}
\end{quote}

By charging more in these situations, repairers earn more in the short term. However, all the data repairers told us that accepting organizational data repair tasks would come with a \textbf{non-disclosure commitment}, maintaining which is an indicator of their \textbf{trustworthiness} that smooths their possibility of getting more of such data repair projects from the same or similar organizations.

\subsection{Collective Protocols for Data Repair Market Sustainability} 
For long-term survival in the data repair market, repairers document and protect their knowledge and avoid sharing it openly. This theme shows the strategies repairers have developed to safeguard their community and sustain it in the long run.

\subsubsection{Restricting Half-Trained Knowledge}
Four lab experts told us that one of their strongest concerns is that half-earned knowledge can be dangerous. If someone learns only a little about the field and then begins practicing on their own, they may accidentally damage a device or corrupt the data completely. When this happens, the entire market suffers. If the data had been intact, at least one of the trained experts could have recovered it, and the community would have benefited. But once the data is destroyed, no one can earn from it, and the customer also loses trust in the entire market. Because of this, repairers carefully check the background of new learners and avoid training those who may not complete the process fully. Lab expert P1 described a case where sharing knowledge casually created exactly this type of problem:

\begin{quote}
\textit{``Once, I casually showed someone a few basic steps about data recovery. He sat beside me for a few days, watching how I worked. Later, he bought his own tools and started calling himself a data recovery expert. One time, I received a hard drive from a woman, and when I spoke with her on the phone, she told me that before coming to me, she had already sought repair from someone else. That repairer turned out to be the same guy I had half-taught. But he completely damaged it by attempting recovery the wrong way. By the time it came to me, I also could not fix it anymore.'' (P1)}
\end{quote}

This example shows how quickly half-trained individuals can harm the reputation and income of the entire community. By preventing incomplete training and keeping knowledge restricted, repairers protect themselves from these risks. For them, secrecy is not only about protecting skills but also about ensuring that the market does not collapse because of inexperienced and unskilled repairers. 

\subsubsection{Limiting Training to Control Market Entry}
Seven lab experts said that they avoid training new people unless they receive a very high fee for it. The main reason is that they see the entry of new repairers as a direct threat to their market. Since the customer base in Bangladesh is already small, every additional person entering the field increases competition and reduces the income of those already working. Repairers, therefore, choose to restrict access to training as a way of keeping the number of experts low and protecting their own position. By limiting who can enter the market, they aim to secure their long-term sustainability in a field that has not yet grown. P17 explained,

\begin{quote}
\textit{``In Bangladesh, data recovery is still not important to most customers. People are not fully dependent on digital storage, since they continue to work on paper documents and keep physical photos. Because the data recovery is expensive and the customer base is small, if another person enters this field, my market value will decrease due to added competition. For these reasons, I do not share my knowledge with others, and I do not want anyone new to enter this work. When everything becomes fully digital, I may think differently, but for now I do not want that.'' (P17)}
\end{quote}

This example shows how protecting knowledge and restricting entry into the market is also a deliberate strategy for long-term survival. Unlike lo-fidelity repair work, the demand for high-fidelity and high-priced data repair is not yet strong enough to support many professionals, as we were told, so they guard their skills carefully and avoid training new competitors. 

\subsubsection{Resilience Against Formal and Professional Competitions}
Five lab experts also told us that even police and CID officers sometimes come to them for help with data repair in criminal investigations. While the officers already have their own data repair teams, they often depend on independent repairers. As P4 explained-

\begin{quote}
\textit{``In this field, the main problem is not the lack of tools or training, it is that when you get stuck, there is no one to help you figure out the next step. Some tasks, such as HDD firmware faults, chip-off dumps, and proprietary DVR video, are very risky because it is often unclear how to proceed safely. Police officers (referring to the technicians from police labs) often get stuck with these cases and become afraid to proceed, since a wrong step can cause irreversible data loss. That is why the officers no longer take the risk and instead bring the devices to us.'' (P4)}
\end{quote}

However, repairers show strong resilience when it comes to sharing their methods. They often do the work of officers but make sure not to disclose their techniques, even if the officers ask them to. For their long-term survival in the market, they keep their knowledge hidden to avoid competition. As P16 said,

\begin{quote}
\textit{``Police and CID officers also ask us about the methods we use, trying to learn how we recover the data. We do not share these processes because if they come to know our techniques, they might start doing the recovery work themselves, and then they would no longer need us. Since they come to us, we earn an income, but if they could do it on their own, they would have no reason to bring the work to us anymore.'' (P16)}
\end{quote}

\newedits{Together, these practices show how repairers protect their knowledge to secure long-term survival. Their resilience ensures that their skills remain indispensable even when facing competition from formal institutions.}

\section{Discussion}
This paper examines Data Repair practices in Dhaka, Bangladesh. Our findings show how repairers sustain in this ecosystem by growing informal expertise, using pirated tools, and developing community-based knowledge. We also noted different moral aspects of Data Repair practices. Our findings contribute to HCI by showing how postcolonial inequities, ethical conflicts, data valuation, and community curation reshape date repair and maintenance. \newedits{These findings closely engage with HCI’s ongoing discourse on postcolonial computing, ethics in big data and AI, repair and maintenance, values in data, and debates around extractivism. By placing data repairers’ everyday practices within these wider scholarly threads, our discussion connects the empirical themes to questions of design, infrastructure, and policy. Our work builds on these connections to offer implications for HCI and to inform policy conversations across both the design and theoretical ends of the field.}

\subsection{Design Implication}
We start by discussing some possible design implications for immediate response to ongoing matters in the data repair space of this community. \textbf{One notable finding} was the heavy reliance on pirated or cracked recovery tools, which exposes repairers to technical risks and legal precarity \newedits{(see 4.4, 5.2.2)}. A value-sensitive design \cite{friedman1996value} response would be to develop lightweight, affordable, and open-source data repair platforms and an alternative stack of resources that recognize the realities of Global South economies while embedding safeguards for privacy, transparency, and intellectual property concerns. One such design response to IP vs. open access discourse would be \textit{Scientific Electronic Library Online (SciELO)}, a Latin American initiative that built an open-access publishing infrastructure to circumvent restrictive academic publishing IP regimes, ensuring equitable access to knowledge without piracy \cite{packer2009scielo, babini2015latin}. We can imagine drawing on their experiences and insights, along with our further field-level engagement, to establish a similar arrangement for local repairers.\textbf{ A second challenge} we noted was the restricted circulation of repair knowledge, where secrecy helps protect livelihoods but also limits broader sustainability \newedits{(see  5.4.1, 5.4.2, 5.4.3)}. Here, design can support community-led archives or cooperative repositories, curated by repairers themselves, that document legacy software, repair scripts, and contextual knowledge without stripping communities of ownership. One such design response that protects restricted knowledge and also allows limited circulation for long-term preservation would be \textit{Mukurtu CMS}: a community-driven content management system originally developed with Australian indigenous communities, that enables ``cultural protocols" so that only certain groups (e.g., elders, high-ranking members of the nation) can access or share knowledge, supporting preservation while respecting boundaries \cite{christen2012does, MukurtuAbout}. We argue for learning from such other marginalized groups' initiatives of protecting their knowledge and practices. By addressing both precarious tools and fragile knowledge infrastructures, design interventions can affirm the dignity of repair work while building more equitable ecosystems for sustaining digital life.

\subsection{Broader Implications for Theory and Policy-making}
Our research connects to HCI by showing how postcolonial inequities, obsolescence, ethical conflicts, data valuation, and community curation reshape concerns with repair and maintenance, while also informing policy through calls for equitable access, legacy preservation, context-sensitive ethics, recognition of marginalized data, and accountable knowledge infrastructures. We discuss them below. 

\subsubsection{Postcolonial Asymmetries in Access to and Practice of Data Repair}
Our research engages with ongoing concerns in postcolonial computing about how repair work in the Global South is shaped by unequal access to knowledge and resources. Scholars have shown that official channels of maintenance, such as manuals, vendor-approved tools, and formal support networks, are often inaccessible to technicians outside the Global North \cite{irani2010postcolonial}. This inaccessibility reflects a deliberate design orientation that centers Western devices, operating systems, and update cycles, while treating Global South contexts as expendable. \newedits{This exclusion is evident in our fieldwork, where repairers routinely encountered barriers when attempting to access sanctioned data recovery solutions (see 5.1.1, 5.1.2)}. The result is a landscape where critical infrastructures for repair are deliberately restricted, reinforcing dependency and inequality. At the same time, data repair practices draw heavily on unauthorized and improvised knowledge. \newedits{Repairers in Dhaka rely on pirated software, cracked dongles, and undocumented methods to restore lost files (see 5.2.2), echoing Sundaram's notion of ``pirate modernity" \cite{sundaram2009pirate} and Karaganis' framing of piracy as survival infrastructure \cite{karaganis2011media}}. These forms of expertise are rarely documented in official or even informal repositories, yet they sustain everyday digital life. \newedits{The very fact that widely used locally manufactured phones remain unsupported by global recovery tools (see 5.1.1) illustrates how millions of Bangladeshi users are rendered technically ``illegible" by design.} We surface a new form of postcolonial inequality that actionize the systematic production of “\textbf{data illegibility},” where postcolonial users and devices are excluded from global data repair and recovery ecosystem, further reinforcing the \textbf{data voids} and \textbf{data poverty} problems. By foregrounding how repairers must navigate between inaccessible official systems and fragile unauthorized tools, we demonstrate that data repair is not only a workaround to scarcity but a site where postcolonial power is actively reproduced and contested through informational maintenance itself. \newedits{Thus, we not only fill HCI's knowledge gap around Global South data poverty, data voids, and risks of future design of biased, erroneous, and culturally non-resonating AI and data-driven systems, we also also expose a politics of survivability through data repair research, and we argue that \textbf{postcolonial HCI design and practice must intervene in the structural conditions that manufacture data illegibility and advance design strategies that do not merely accommodate Global South users but actively redistribute the power to maintain, preserve, and repair the digital traces that shape their futures.}}

\subsubsection{Against Obsolescence: Preserving Repair Knowledge}
Our research also engages with debates in HCI and sustainability studies that critique the culture of constant updating and replacement in the technology industry. Scholars such as Jackson and Kang \cite{jackson2014breakdown} and Dourish \cite{dourish2010hci} have argued that obsolescence is built into design logics that privilege novelty over longevity. This orientation undermines repair, particularly in the Global South where users depend on older hardware and outdated software to sustain access. \newedits{Our study showed how repairers face difficulties locating drivers, protocols, and tools that have been abandoned by manufacturers but remain necessary for recovering valuable data (see 5.1.2)}. Preserving old mechanisms and local repair knowledge is essential to prevent the erasure of digital lives tethered to outdated formats. Lepawsky and Mather \cite{lepawsky2011beginnings} highlight how waste materials from the North are repurposed in the South, revealing the hidden sustainability of reuse economies. \newedits{Similarly, data repair requires access to legacy software and maintenance protocols to restore fragile archives (see 5.1.2)}. By examining these practices, our research argues for preservation infrastructures that treat old repair knowledge as a global resource essential for sustaining equity in digital access, rather than as expendable remnants of technological progress. \newedits{We contend that preserving repair knowledge is essential for interrupting the epistemic erasures built into global technology cycles, and that treating this knowledge as a shared infrastructural resource calls on HCI to design concrete mechanisms, policies, and platforms that actively resist the marginalization of Global South users in future AI- and data-driven ecosystems.}

\subsubsection{Repair Ethics in the Shadow of AI}
Our research engages with ongoing discussions about the ethical dimensions of repair, particularly in contexts where sensitive personal data is at stake. Nissenbaum's theory of contextual integrity \cite{nissenbaum2004privacy} emphasizes that privacy is culturally situated, and ICTD studies show that repairers often navigate complex ethical expectations without institutional safeguards \cite{ahmed2015learning,houston2013inventive}. \newedits{In Dhaka, repairers cultivate trust and discretion while handling private archives, balancing professional survival with community accountability (see 5.3.1, 5.3.3, 5.4.1, 5.4.3)}. However, these situated ethics are increasingly challenged by global technologies such as large language models. \newedits{When consulted, LLMs often refuse to provide guidance for tasks involving cracked tools or encrypted data, citing universal intellectual property norms (see 5.2.2)}. This stance overrides the contextual ethics developed by repairers and undermines the practical needs of their communities. Similar tensions have been observed in other Global South settings, where local practices are delegitimized by global standards \cite{jackson2012repair}. \newedits{To address these conflicts, we suggest that OpenAI and similar tech companies be more transparent about how regional and international intellectual property considerations shape their data practices. Such transparency and more equitable access to information are essential for repairers who rely on these tools in constrained environments. While current systems often respond restrictively in cases involving cracked tools, restricted formats, or encrypted data, global technologies should not only limit access but also explore approaches that liberate regional policy contexts. They must also remain sensitive to the needs of the Global South while still respecting legal and ethical boundaries. By highlighting these conflicts, our research argues that data repair exposes the limitations of universalist ethical frameworks and calls for plural approaches that respect and integrate local and regional values and sensitivities of responsibility.}

\subsubsection{Repairing Inequities in Data Valuation}
Our work also speaks to concerns about the uneven valuation of data in global digital economies. HCI and HCI4D scholarship have shown that the digital practices of marginalized users are often overlooked or treated as peripheral \cite{wyche2015exploring,ahmed2015learning}. Official service centers frequently dismiss requests to restore low-income users' personal files, framing them as insignificant compared to corporate or institutional data. \newedits{Our study documents how local repairers instead recognize the emotional and cultural weight of these materials, treating wedding photos, memorial videos, or educational files with exceptional care (see 5.3.1, 5.3.3)}. \newedits{Our findings show that emotional valuation is closely tied to customers’ digital literacy and vulnerability. Repairers noted that clients who are less comfortable with technology often receive special care, while also becoming more dependent on repairers due to their limited technical knowledge (see 5.3.1)}. At the same time, \newedits{repairers strategically use emotional value in their pricing decisions—for example, charging more when they know a customer is deeply attached to the data (see 5.3.1)}. These situations create social ties that are long-term yet often insensitive and error-prone, reinforcing customer dependence on repairers. Together, these dynamics illustrate how pricing practices structure equity, trust, and dependency within the repair ecosystem. \newedits{These pricing practices, including loyalty discounts (see 5.3.2), shape long-term trust, return behavior, and informal customer–provider relationships, which are central concerns in HCI’s study of engagement and service design in informal markets.} Jackson and Kang \cite{jackson2014breakdown} remind us that repair is motivated by love, memory, and responsibility as much as by utility. \newedits{Data repair makes these values tangible, especially for communities whose data is often devalued (see 5.3.1, 5.3.2, 5.3.3)}. \newedits{By revealing how emotional, social, and cultural valuations of data exceed the narrow economic logics embedded in global systems, our research argues that HCI must confront the structural inequities that render poor people's data disposable and design infrastructures that treat all data as socially significant and worthy of preservation. Therefore, HCI researchers need to develop frameworks and and preservation infrastructures that recognize these lived meanings rather than reproducing hierarchies that privilege corporate and institutional data.}

\subsubsection{Community Stewardship Against Extractivism}
Finally, our research engages with discussions of knowledge curation and community capital in repair practices. Lave and Wenger's work on communities of practice \cite{lave1991situated} and Ahmed et al.'s ethnography of Dhaka repairers \cite{ahmed2015learning} show how expertise is transmitted through apprenticeship, secrecy, and selective access. \newedits{Our findings reveal that repairers curate and restrict their knowledge not only to maintain economic survival but also to build trust and reputation within their communities (see 5.4.1, 5.4.2, 5.4.3)}. At the same time, there is a pressing need to preserve and amplify these repair knowledges at a broader scale. Rosner and Ames \cite{rosner2014designing} suggest that infrastructures of breakdown can be leveraged as sites of innovation, but only if knowledge is not extracted without reciprocity. Postcolonial critiques warn against extractivism, where local knowledge is removed from its context and commodified for external benefit \cite{irani2010postcolonial}. By situating repair knowledge as both a local asset and a global necessity, we argue for accountable infrastructures that preserve repair knowledges while strengthening the repair communities of the Global South, ensuring they are recognized not as peripheral labor but as central stewards of equitable data futures. \finalrev{Even this manuscript that reports a part of our broader project might not have captured our positionality regarding extractivism fully;  that extensive engagement with repair communities and other stakeholders through long-term ethnography and codesign is core to establishing an accountable data repair ecosystem.} \newedits{This would require HCI and local governments to join forces to institutionalize community-centered data and AI curation and repair practices, ensuring more equitable and representative data-driven futures.} 

\subsection{Limitations and Future Work}
Our research has several limitations. First, our ethnography was conducted in Dhaka. While this provides rich depth, the findings cannot fully capture the heterogeneity of data repair practices across Bangladesh or other Global South contexts. Second, we deliberately did not access the actual content of repaired data to protect privacy. While ethically essential, this limited our ability to analyze how different data types may shape repair strategies. Third, our focus on repairers' perspectives means that we foreground their voices more than those of customers, policymakers, or tool developers. This lens highlights the lived realities of repairers but leaves complementary viewpoints underexplored.

Future work can broaden the scope of data repair research in two ways. First, cross-contextual studies across different Global South regions are needed to understand how cultural, infrastructural, and economic variations shape repair ecosystems, and to identify both shared struggles and unique adaptations. Second, future research should involve prototyping design and policy interventions, such as open-source and community-led platforms for data repair, to test how value-sensitive design can reduce reliance on piracy while preserving local autonomy and knowledge. \finalrev{Third, to build accountable infrastructure and better moderate data for the local community, our future work focuses on identifying and setting benchmarks through ethnography and co-design with local repair communities to ensure alignment with community values.}. Together, these directions can extend the theoretical and practical impact of this work, building toward a more equitable and sustainable global data repair ecosystem.

\section{Conclusion}
This paper introduced data repair as a critical but overlooked dimension of HCI, grounded in an ethnographic study of Dhaka's repair ecosystem. By documenting the interplay of restricted knowledge, pirated tools, ethical tensions, and the social valuation of data, we show that repairers' practices both sustain fragile digital life and reveal the inequities of global technological infrastructures. Our analysis advances HCI theory by extending repair scholarship into the informational domain and highlights data repair as a site of postcolonial struggle, market survival, and community curation. For design and policy, we call for infrastructures that support equitable access, preserve legacy knowledge, respect local ethics, and recognize the value of marginalized people’s data in building just data futures.

\bibliographystyle{ACM-Reference-Format}
\bibliography{sample-base}

\end{document}